# DEVELOPING ENTERPRISE CYBER SITUATIONAL AWARENESS

Christopher L. Gorham

Capitol Technology University, USA

*ABSTRACT*

*This paper will examine why its become important for organizations to develop an comprehensive Information Technology (IT) Modernization strategy that focuses on improving their business process and protecting their network infrastructure by leveraging modern-day cybersecurity tools. Most of the topic will focus on the U.S. Department of Defense's (DOD) strategy towards improving their network security defenses for the department and the steps they've taken at the agency level where components under DOD such as DISA (The Defense Information Systems Agency) are working towards adding tools that provides additional capabilities in the cyber space. This approach will be analyzed to determine if DOD goals address any of their vulnerabilities towards protecting their networks. One of the agencies under the DOD umbrella called DISA (The Defense Information Systems Agency) provides DOD a template on how to build a network that relies upon layers of security to help it combat cyber attacks against its network. Whether that provides an effective solution to DOD remains a question due to the many components that operate under its direction. Managing these networks is the principle responsibilities for the Department of Defense. Nevertheless, it does demonstrates that there are tools available to help DOD build an strong enterprise cyber network of situational awareness that strengthens the ability to protect their network infrastructure.*

## 1. INTRODUCTION

It's generally acceptable that the strategic objective of any IT modernization plan constructed by the U.S. Department of Defense (DOD) should allow itself to take advantage of the benefits that enhances it's ability to innovate. DOD is one of the largest federal agencies in the U.S. federal government. The organizational structure of the DOD was created as the successor to the National Military Establishment in 1947 under the National Security Act of 1947 (50 U.S.C. 401). In 1949, DOD was established as an executive department of the U.S. Federal Government by the National Security Act Amendments of 1949 under code 5 U.S.C. 101. The Department of Defense exist as an organization under the President, in his role as the Commander in Chief and is lead by the Secretary of Defense who provides direction and control over matters that includes military functions for the Army, Navy, Air Force and the Joint Chiefs of Staff. These DOD components provides operational military advice under the commands for various defense agencies that were established for specific purposes.

James Proctor writes in his blog for the Inteq Group, that one of the "overarching benefits of IT modernization is providing the platform to transform business processes that enable the execution of a business strategy" (Proctor, 2019). Mr. Proctor suggests that this allows an organization to become effective in creating a value for it's customer as well as efficiencies for the business overall (Proctor, 2019). When developing a strategy towards leveraging the benefits of IT modernization, organizations need to consider how to improve their process.

It's critically important that organizations such as DOD should develop an Information Technology (IT) Modernization strategy that focuses on three critical elements towards improving their business process and cybersecurity:

1) Consolidating Infrastructure
2) Streamlining Processes
3) Strengthening the Workforce

These elements are strategic goals to support organizations towards achieving their mission objectives by leveraging technology to gauge their effectiveness. The success of the long-term approach towards this strategy is dependent on the success of their short-term approach, which requires determining their current and future needs. Businesses will need to develop a partnership with the tech community to determine how to improve business process while measuring their effectiveness and user satisfaction. This also includes how to reduce costs and wasteful spending while improving cyber security and interoperability methods. Ultimately, the goal for any organization that is focusing on an agile, fast and responsive to the delivery of their IT capabilities that is robust and maintains the highest level of cyber security.

As technology becomes more centric in the lives of people in the 21st century, organizations are developing their business strategies around being efficient at delivery services or products in the hands of consumers. But with being efficient leads to more business processes becoming automated. Automation is the driving force behind business information systems producing data that allows organizations to learn more about their customers, consumer or users. It allows them to gain an understanding of the data through metrics that measures the performances of their process internally and externally. However, the quantity of data being produced by information systems can grow at an increasingly accelerated rate if not measured properly. Redundant data is often one of the results in managing data when different systems have to communicate with each other. This can sometimes lead to data redundancies where one system is collecting the same data it stores in its own database. To reduce this inefficiency, information technology (IT) professionals must establish a process that reduces redundancies and inefficiencies from old legacy systems while developing new and modernized software applications that opens the door to more effective automation. For this task to be accomplished, organizations need to address the principal challenges they face in determining what areas of their system will need automation and also determining the workforce ability in meeting their automation objectives.

Automation presents many challenges to overcome to implement successfully. This research will focus on some of the challenges organizations will need to address if they are to mitigate to be efficient in managing their business process. Those challenges include analyzing what areas need automation, the relationship between the client, business user and IT department, the different roles in IT that are needed to build a successful system and developing an end to end process that keeps the system updated from cyber attacks. All these principal challenges will be reduce or minimize from the standpoint of addressing inefficiencies of the interfaces between enterprise application systems.

## 2. TECHNOLOGY AND INFORMATION AWARENESS

One of the benefits for DOD's moving to the cloud is the opportunity to leverage the services that will enhance their capabilities such as managing data and analytics. DISA's solution may

International Journal of Managing Information Technology (IJMIT) Vol.12, No.3, August 2020

offer a framework for other organizations outside of DOD that takes advantages of innovative cloud solutions tools to protect their network infrastructure from cyber attacks. This is yet to be establish and will be incorporated into the review and analysis.

If DOD is committed towards building a modernized cyber defense system, then it must consider the risks associated with such a strategy. This may include the difficulty of migrating from legacy systems that many of it's components rely upon towards managing their data on the current network. However, there are other negative factors that come into play that may hinder efforts for DOD to modernize their IT networks.

## 3. DOD's Digital Enterprise Strategy

DOD has developed their Digital Modernization strategy that presents it's goals and objectives while aligning their digital priorities in presenting their central vision of creating a secure, coordinated, transparent, and cost-effective IT architecture (Barnett, 2019). This strategy will allow DOD to transform data into actionable information and ensure dependable mission critical execution while dealing with the persistent threat of cyber warfare (Barnett, 2019).

Some of the top priorities that stands out from DOD's digital modernization strategy are: (1) Cybersecurity; (2) Artificial Intelligence (AI); (3) Cloud; and (4) Command, Control, and Communications (C3) (Barnett, 2019). These priorities will help guide the mission towards advancing mission critical goals that sets forth in the strategy in providing (1) Innovate for Competitive Advantage; (2) Optimize for Efficiencies and Improved Capability; (3) Evolve Cybersecurity for an agile and Resilient Defense Posture; (4) Cultivate Talent for a Ready Digital Workforce (Barnett, 2019).

DOD hopes their Modernization Strategy will align with their high-level goals to meet the priorities of the agency's leadership. The strategy will highlight the collaboration among DOD, their industry partners, and non-DOD business partners (Barnett, 2019). This will help DOD leadership understand which IT investments should focus on how to enable and create better solutions for the agency. The strategy will represent a shift from an organization standpoint that will allow the DOD Chief Information Officer (CIO) to oversee IT budget requests for modernization efforts while developing within the framework of a comprehensive management system(Barnett, 2019).

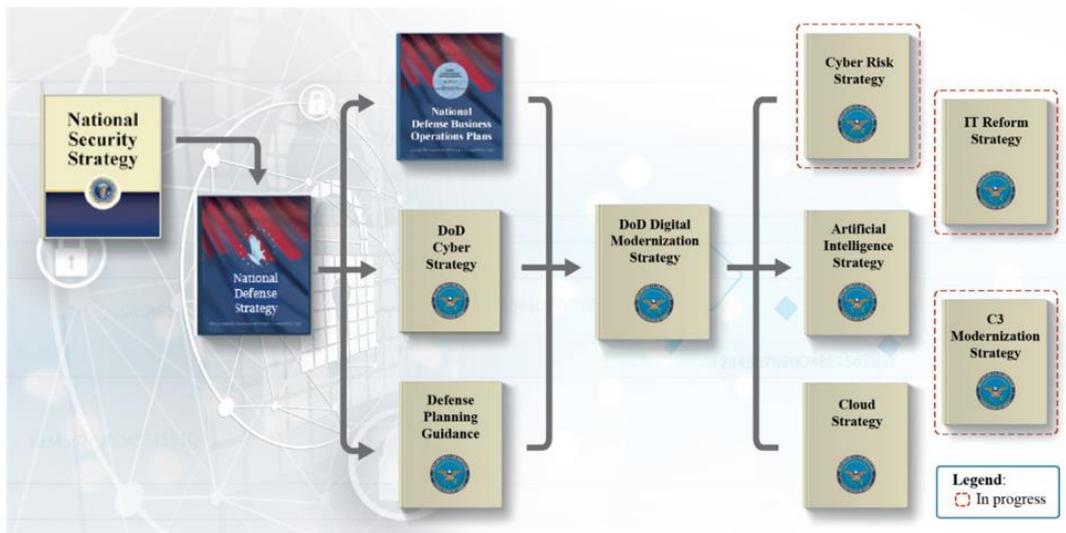

Figure 1. Source: DoD Digital Modernization Strategy: FY 19-13

## 4. ARMY PLANS TO FOLLOW NAVY, AIR FORCE IN OUTSOURCING MUCH OF ITS IT INFRASTRUCTURE

One of the Department of Defense's components, the Army, will soon undertake another strategy to modernize their IT infrastructure into the next generation. The Army, along with the Navy, will invest in the Enterprise-IT –as-a-Service platform for the purpose of consolidating their IT infrastructure and reducing the workforce head and costs. This approach follows DOD's strategy of modernizing their technology to improve their cyber defenses but by having so many components at the agency level with their own infrastructure and IT workforce, a multi-phased approach may offer the best option. Currently, the Army has an estimate of 70 percent of their routers, servers and devices for end-users located in 288 world-wide facilities across the world (Serbu, 2019). That number is close to 90 percent for voice communications, according to an article written by Jared Sarbu titled "The Army plans to follow Navy, Air Force in outsourcing much of it's IT infrastructure". The Army's Chief Information Officer (CIO), LT. Gen. Bruce Crawford, estimates that on the current track of replacing all of the gear the Army has stored for it's IT infrastructure will take until 2030 under the current modernization methodologies (Serbu, 2019).

A new approach was considered and adopted where the Army would instead issue an RFI (Request For Information) for vendors to restructure their IT infrastructure in a quicker way that speeds up their modernization efforts (Serbu, 2019). The goal would be for the Army to plan out a "three pilot projects to test out the IT-as-a-service" model in 2019 to which the infrastructure will be operated by contractors instead of government personnel (Serbu, 2019). The objective towards this approach if successful will be about delivering more effective IT services across the Army organization and serve as a model for success for DOD as a whole as they migrate their IT network to the cloud.

## 5. DEVELOPING A STRATEGY FOR CYBERSECURITY IN THE CLOUD

DOD has constructed a plan towards consolidating all of their networks into an eco-system that should be able to provide a more efficient way of delivering data. This effort begins with reducing their data centers from 770 to less than 100(Insinna, 2013). That helps to reduce cost while maintaining high-level performance for all remaining networks. DOD will also consolidate their Network Operating Centers from 65 to 25 while transitioning to a Joint Enterprise Architecture via secure access for their employees (Insinna, 2013).

To migrate to an enterprise cloud infrastructure, DOD will need to develop a plan to execute a strategy that streamlines standards for a cyber-secure cloud environment. This strategy will only leverage commercial cloud platforms that meet DOD's cyber security requirements. This will help DOD lay down the foundation of automation of many repetitive tasks that slows down the workforce. Many of the military technical tasks can be targeted for automation but because of additional cost measures, implementing this task can be burdensome. To address these liabilities, DOD must develop a strategic plan that will meet the needs for DOD to standardize their IT platform.

In an effort to standardize their IT platforms, DOD will develop a platforms that centers around supporting a cloud-secure environment. This will help them to use a unique standard for the platforms that leverages applications that stores data around a Common Platform. The "Common" platform consist of PaaS (Platform-As-A-Service), messaging and identity management. Together, DOD will be able to build a Common Infrastructure that transports, facilitates and processing data and files in secure platform. This effort will build the infrastructural foundation towards innovation for the DOD workforce. But some realities in regards to the effect of these efforts to modernize their network will most likely lead to a discussion on determining if automation if a benefit or consequence for the workforce.

## 6. THE EFFECT OF AUTOMATION ON THE WORKFORCE

In an article written by James Bessen titled "Automation Reaction – What Happens To Workers At Firms That Automate", he states that the impact of automation on workers will most likely have an effect on workers when firms increase the probability of workers separating from their employers and decreases the days they worked (James E. Bessen, 2019). Mr. Bessen based his conclusion off measuring automation at firms that studies the worker's impacts of automation from how it originated. Mr. Bessen worked with a research group to develop a methodology that combines event level study with analysis while leveraging the timing of "firm-level automation" to identify the effects on the workforce (James E. Bessen, 2019). He then focuses on the automation events when they occur across the non-financial sectors of the economy in order to identify a specific automation technology in isolation of other events (James E. Bessen, 2019).

This allows him to measure an assortment of outcomes that identify workers who've stayed employed at the firm during the years surrounding the automation event. He found that automation increases the probability of workers leaving their employers which for those who stayed at least three years longer versus the separation of employment decreased in annual days worked. The effect led to a "5-year cumulative wage income loss" resulting of 8% of one year's earnings (James E. Bessen, 2019). The results from wage income losses were disproportionately impacted by more senior workers and partially affected depending upon the benefit system of the employer. Bessen's research study on automation focused on comparing their findings from a literature on mass layoffs that provided more detailed results of the effects of automation from a gradual standpoint to determine if automation displaces fewer workers than projected from other studies conducted (James E. Bessen, 2019). While his findings don't predict the "apocalyptic" outcome of mass unemployment it does show that automation will somewhat have a negative impact on workers depending upon their type of work along with their seniority status at the firm they are employed (James E. Bessen, 2019). Based off those observations, it would be difficult to determine on a widespread scale the outcome on workers as automation increases across the economy.

## 7. THE GENERATIONAL SHIFT FOR IT PROFESSIONALS

As a result of technology's impact on automation for the workforce, the way people perform their jobs will have a dramatic effect on the skills required to obtain the next-generation of jobs. In an article titled "The Generational Shift in IT Drives Change for IT Pros,", Steve Cox suggests that the emergence of the cloud and other technologies have enables forces to bring about change to organizations and the IT workforce (Cox, 2018). The impact of cloud technologies have given businesses a bigger role in determining the future of application delivery (Cox, 2018). IT professionals also are partners with businesses because they play a key role in using "technology-driven creativity to enable innovation" while simplifying business processes for the organization. The IT industry is moving fast towards adopting emerging technologies such as artificial intelligence, blockchain and the Internet of Things (IoT) that are built within the cloud platform such as Amazon Web Services (AWS) and Microsoft's Azure technologies (Cox, 2018). In order for businesses, governments such as DOD and the IT workforce in general to move forward, they must be partners in helping their stakeholders develop an innovate approach in building newer applications. This will require a level of understanding from all parties that the way work being is defined will change in the future in addition to the skillsets necessary for workers to perform their tasks. IT professionals specifically have to perform the critical part by staying in the loop on the latest technologies (Cox, 2018). Embracing this approach will provide the flexibilities for organizations and government agencies such as DOD to respond to challenges in cyber warfare more quickly while maintaining their innovative approach in adopting emerging technologies.

## 8. ANALYSIS

DOD's Air Force CIO Dana Deasy spoke about the agency's modernization efforts at a IT Day event in Washington last year (Vergun, 2019). Deasy spoke about how the digital modernization efforts focuses on making network improvements, control and communications by leveraging cloud computing and artificial intelligence (Vergun, 2019).Deasy also mentioned DOD's efforts towards building out the Joint Enterprise Defense Infrastructure environment which is it's cloud computing platform that will be available for all DOD agencies offering all cloud-based services (Vergun, 2019).

One of the main priorities of the JEDI contract will be efforts in building our DOD's 5G network. DOD plans to work with NATO to integrate digital modernization planning with partners that will develop proposals for 5G network dynamic spectrum sharing and start hosting 5G network pilot programs at selected installations, according to Deasy (Vergun, 2019).Other milestones would include the development of performance standards for supply-chain risk. This will begin by issuing a cyber workforce management strategy to ensure it's compliance for 28 of 30 cyber hygiene metrics to keep it's data safe (Vergun, 2019).

Whether the efforts by DOD is successful or not will depend on investments made in cybersecurity. The federal government shutdown in January 2019 actually represented a security risk since funding stalled during dispute and prevented any important work from being completed. Phil Goldstein wrote in a FedTech article titled "How Did the Government Shutdown Affect Federal Cybersecurity", that critical cyber initiatives for security were put on hold during the gov't shutdown (Goldstein, 2019). As a result, the Department of Homeland Security's (DHS) Cyber Security Office, which only became operational in November, wasn't able to get operations fully up to date and had a difficult time hiring cybersecurity workers that needed to fill critical roles (Goldstein, 2019).

This event demonstrated that the U.S. federal government's budgeting process itself could cause problems for any federal agency looking to modernize their IT infrastructure. The process is highly complicated due to the political process that often causes a stalemate. The President and Congress have to be in agreement on the funding necessary for DOD to modernize their network infrastructure. If the project isn't funded in a way that properly addresses all vulnerable areas, DOD risks their network becoming increasingly susceptible to cyber attacks, which could harm their networks. The manpower necessary to manage the network could be costly due to the required skillsets needed to modernize and future proof or encrypting their data against cyber attacks. Developing a strategic plan that centers on outsourcing their infrastructure to the private sector so their networks can help leverage modern tools to rebuild their network for the next generation. This will provide DOD a strategic path forward in combating $21^{st}$ century cyber threats against their network while maintaining a foundation for the future.

Another area of concern is how much of a factor is automation going to be for the workforce of DOD. The Army's approach in Serbu's article takes a modernization approach in re-architecturing their network for the future(Serbu, 2019). As Serbu stated, "the Army would instead rely upon an RFI (Request For Information) for vendors to restructure their IT infrastructure in a quicker way that speeds up their modernization efforts (Serbu, 2019)." In outsourcing their efforts to quickly adopt their modernization efforts, the effects on the workforce could be a major concern for IT workers looking to maintain and update their skills. New technology forces workers to adopt a new approach in keeping up with the ever growing changes in the IT field.

## CONCLUSION

DOD will need to analyze this shift closely to determine if the positions that currently exist in their workforce meet the demand of the skills necessary to modernize their infrastructure. This will eventually lead to a discussion of the quality of training DOD is willing to provide for their workers as well as incentives to attract and maintain the talent level of their IT workforce. These kinds of efforts by any organization will always resolve one problem area but create a new one in a cause-and-effect type of scenario. DOD will need to confront these challenges while maintaining an approach that balances their efforts to modernize their network and adopting new methods towards building a modern IT workforce for the future.